\documentclass[prd,twocolumn,groupedaddress,showpacs,nofootinbib]{revtex4}
\usepackage{graphicx}
\usepackage{dcolumn}
\usepackage{amssymb}
\usepackage{mathrsfs}
\usepackage{amsmath}
\usepackage{epsfig}
\usepackage[dvips]{color}
\usepackage{hhline}

\begin{document}

\title{Mirror symmetry: from active and sterile neutrino masses
to baryonic and dark matter asymmetries}

\author{Pei-Hong Gu}
\email{peihong.gu@mpi-hd.mpg.de}

\affiliation{Max-Planck-Institut f\"{u}r Kernphysik, Saupfercheckweg
1, 69117 Heidelberg, Germany}

\begin{abstract}

We consider an $SU(3)'^{}_c\times SU(2)'^{}_L\times U(1)'^{}_Y$
mirror sector where the field content and dimensionless couplings
are a copy of the $SU(3)^{}_c\times SU(2)^{}_L\times U(1)^{}_Y$
ordinary sector. Our model also contains three gauge-singlet
fermions with heavy Majorana masses and an $[SU(2)^{}_L\times
SU(2)'^{}_L]$-bidoublet Higgs scalar with seesaw-suppressed vacuum
expectation value. The mirror sterile neutrino masses will have a
form of canonical seesaw while the ordinary active neutrino masses
will have a form of double and linear seesaw. In this canonical and
double-linear seesaw scenario, we can expect one sterile neutrino at
the eV scale and the other two above the MeV scale to fit the
cosmological and short baseline neutrino oscillation data.
Associated with the $SU(2)^{}_L$ and $SU(2)'^{}_L$ sphaleron
processes, the decays of the fermion singlets can simultaneously
generate a lepton asymmetry in the $[SU(2)^{}_L]$-doublet leptons
and an equal lepton asymmetry in the $[SU(2)'^{}_L]$-doublet leptons
to explain the existence of baryonic and dark matter. The lightest
mirror baryon then should have a determined mass around
$5\,\textrm{GeV}$ to account for the dark matter relic density. The
$U(1)$ kinetic mixing can open a window for dark matter direct
detection.

\end{abstract}


\maketitle

\section{Introduction}

Various neutrino oscillation experiments have firmly established the
three active neutrino oscillation picture \cite{ftv2012}, however,
some short baseline neutrino oscillation experiments
\cite{aguilar-arevalo2001} and recent re-evaluations of reactor
antineutrino fluxes \cite{mueller2011} hint the existence of
additional sterile neutrinos with the eV-scale masses
\cite{kms2011,afghm2013,kmms2013}. The light active neutrinos can be
elegantly understood in the seesaw
\cite{minkowski1977,mw1980,flhj1989} or other
\cite{zee1980,mahanta1999,knt2002,gs2008,koy2013,rw1983,gs2007,mohapatra1986,barr2003,mv1986,fp2009,gu2011}
extensions of the $SU(3)^{}_c\times SU(2)^{}_L\times U(1)^{}_Y$
standard model (SM). If the light sterile neutrinos are eventually
confirmed, we need explain not only the existence and small masses
of the sterile neutrinos but also the mixing between the active and
sterile neutrinos. Such sterile neutrinos can naturally appear in
the mirror
\cite{ly1956,kop1966,pavsic1974,bk1982,glashow1986,flv1991,hodges1993,silagadze1997,cf1998,
mt1999,bcv2001,bgg2001,iv2003,fv2003,foot2004,acmy2009,dlnt2011,chly2012,gu2012-2,foot2012,flv1992,abs1992,bdm1996,bb2001,bb2001-2,berezhiani2004,berezhiani2005,berezhiani2006,
bb2006-1,bb2006} universe models discussed in the literature
\cite{flv1992,abs1992,bdm1996,bb2001,bb2001-2,berezhiani2004,berezhiani2005,berezhiani2006,
bb2006-1,bb2006}. There are also other ideas for the light sterile
neutrinos
\cite{kty2010,mohapatra1991,gu2010,mn2011,gh2011,ct2011,abazajian2012}.
On the other hand, precise cosmology has indicated that in the
present universe the energy density of the dark matter is comparable
with that of the ordinary matter \cite{komatsu2010}. This raises an
interesting possibility that the dark and ordinary matter may have a
special relation although their properties are very different. For
example, the dark matter relic density may be an asymmetry between
the dark matter and antimatter
\cite{flv1992,abs1992,bdm1996,bb2001,bb2001-2,berezhiani2004,berezhiani2005,berezhiani2006,
bb2006-1,bb2006,ly1956,kop1966,pavsic1974,bk1982,glashow1986,flv1991,hodges1993,silagadze1997,cf1998,
mt1999,bcv2001,bgg2001,iv2003,fv2003,foot2004,acmy2009,dlnt2011,chly2012,gu2012-2,foot2012,nussinov1985,bcf1990,kaplan1992,dgw1992,kuzmin1998,kl2005,as2005,
clt2005,gsz2009,dmst2010,bdfr2011,mcdonald2011,hmw2010,dk2011,
frv2011,hms2011,kllly2011,gsv2011,fss2011,myz2012,idc2011,bpsv2011,cr2011,
as2012,ms2011,dm2012,fnp2012,bp2011,ptv2011,ams2012,kv2012,ccs2012,bfk2013,bmp2013}
since the ordinary matter exists as a baryon asymmetry. In
particular, the asymmetric dark matter can be well motivated in the
mirror universe models
\cite{flv1992,abs1992,bdm1996,bb2001,bb2001-2,berezhiani2004,berezhiani2005,berezhiani2006,
bb2006-1,bb2006,ly1956,kop1966,pavsic1974,bk1982,glashow1986,flv1991,hodges1993,silagadze1997,cf1998,
mt1999,bcv2001,bgg2001,iv2003,fv2003,foot2004,acmy2009,dlnt2011,chly2012,gu2012-2,foot2012}.

In this paper we shall propose a novel mirror universe model to give
the active and sterile neutrino masses as well as the baryonic and
dark matter asymmetries. In addition to the $SU(3)^{}_c\times
SU(2)^{}_L\times U(1)^{}_Y$ ordinary sector and its
$SU(3)'^{}_c\times SU(2)'^{}_L\times U(1)'^{}_Y$ mirror partner, our
model contains three gauge-singlet Majorana fermions and an
$[SU(2)^{}_L\times SU(2)'^{}_L]$-bidoublet Higgs scalar. The mirror
symmetry is softly broken to allow the symmetry breaking pattern in
the mirror sector different from that in the ordinary sector. The
mirror photon can become massive according to the mirror
electromagnetic symmetry breaking. By integrating out the fermion
singlets and the Higgs bidoublet, we can get the mirror sterile
neutrino masses by canonical \cite{minkowski1977} seesaw, as well as
the ordinary active neutrino masses by double \cite{mohapatra1986}
and linear \cite{barr2003} seesaw. In this canonical and
double-linear seesaw scenario, two sterile neutrinos are above the
MeV scale and the other one is at the eV scale, so that their
existence can fulfill the cosmological and short baseline neutrino
oscillation data \cite{afghm2013}. Our model can realize a
leptogensis
\cite{fy1986,mz1992,fps1995,hms2000,di2002,hs2004,hlnps2004,hrs2005,dnn2008,bd2009,gu2012}
as a common origin of the ordinary and dark matter. Specifically,
the decays of the fermion singlets can simultaneously generate a
desired lepton asymmetry in the $[SU(2)^{}_L]$-doublet leptons and
an equal lepton asymmetry in the $[SU(2)'^{}_L]$-doublet leptons
even if we do not resort to the fine tuned resonant effect
\cite{fps1995}. Through the sphaleron-induced lepton-to-baryon
conversion \cite{krs1985}, we can obtain an ordinary baryon
asymmetry and an equal mirror baryon asymmetry to account for the
number densities of ordinary and dark matter. The lightest mirror
baryon then should have a determined mass around $5\,\textrm{GeV}$
to explain the ratio between the ordinary and dark matter energy
densities. In the presence of a $U(1)$ kinetic mixing, the dark
matter particle can be verified in the ongoing and future dark
matter direct detection experiments.

Our model shares some ideas of Ref. \cite{acmy2009}, where the
authors introduced two $[SU(2)^{}]$-triplet Higgs scalars to
generate the active neutrino masses by type-II \cite{mw1980} and
inverse \cite{mv1986} seesaw. They also assumed all of the mirror
neutrinos above the MeV scale. Furthermore, they resonantly enhanced
the CP asymmetries in the decays of the fermion singlets to produce
the required lepton and baryon asymmetries.

\section{The model}

There are two Higgs scalars in both of the ordinary and dark
sectors,
\begin{eqnarray}
\label{dhiggs} \phi^{}_{d}(\textbf{1},\textbf{2},+1)
=\!\!\left[\begin{array}{r}\phi^{+}_{d}\\
[1mm]
\phi^{0}_{d}\end{array}\right]\!\!&\leftrightarrow&\!\!\phi'^{}_d(\textbf{1},\textbf{2},+1)
=\!\!\left[\begin{array}{r}\phi'^{+}_{d}\\
[1mm] \phi'^{0}_{d}\end{array}\right]\,,\nonumber\\
[2mm] \phi^{}_{u}(\textbf{1},\textbf{2},-1)
=\!\!\left[\begin{array}{r}\phi^{0}_{u}\\
[1mm]
\phi^{-}_{u}\end{array}\right]\!\!&\leftrightarrow&\!\!\phi'^{}_u(\textbf{1},\textbf{2},-1)
=\!\!\left[\begin{array}{r}\phi'^{0}_{u}\\
[1mm] \phi'^{-}_{u}\end{array}\right]\,.
\end{eqnarray}
Here and thereafter the mirror fields are denoted by a prime on a
symbol, and hence the parentheses following the ordinary fields are
the quantum numbers under the $G=SU(3)^{}_c\times SU(2)^{}_L\times
U(1)^{}_Y$ gauge group, while the parentheses following the mirror
fields are the quantum numbers under the $G'=SU(3)'^{}_c\times
SU(2)'^{}_L\times U(1)'^{}_Y$ gauge group. Our model also contains
three families of ordinary and mirror fermions,
\begin{eqnarray}
\label{fermion} q_{L}^{}(\textbf{3},\textbf{2},+\frac{1}{3})
=\!\!\left[\begin{array}{r}u^{}_{L}\\
[1mm]
d^{}_{L}\end{array}\right]\!\!&\leftrightarrow&\!\!q'^{}_{L}(\textbf{3},\textbf{2},+\frac{1}{3})
=\!\!\left[\begin{array}{r}u'^{}_{L}\\
[1mm] d'^{}_{L}\end{array}\right]\,,\nonumber\\
[2mm]
d^{}_{R}(\textbf{3},\textbf{1},-\frac{2}{3})\!\!&\leftrightarrow&
\!\!d'^{}_{R}(\textbf{3},\textbf{1},-\frac{2}{3})\,,\nonumber\\
[2mm]
u^{}_{R}(\textbf{3},\textbf{1},+\frac{4}{3})\!\!&\leftrightarrow&
\!\!u'^{}_{R}(\textbf{3},\textbf{1},+\frac{4}{3})\,,\nonumber\\
[2mm] l_{L}^{}(\textbf{1},\textbf{2},-1)
=\!\!\left[\begin{array}{r}\nu^{}_{Li}\\
[1mm]
e^{}_{L}\end{array}\right]\!\!&\leftrightarrow&\!\!l'^{}_{L}(\textbf{1},\textbf{2},-1)
=\!\!\left[\begin{array}{r}\nu'^{}_{Li}\\
[1mm] e'^{}_{Li}\end{array}\right]\,,\nonumber\\
[2mm]
e^{}_{R}(\textbf{1},\textbf{1},-2)\!\!&\leftrightarrow&\!\!e'^{}_{R}(\textbf{1},\textbf{1},-2)\,,
\end{eqnarray}
with the family indices being omitted for simplicity. We further
introduce three gauge-singlet fermions \cite{flv1992,abs1992},
\begin{eqnarray}
\label{singlet}
N^{}_{Ri}(\textbf{1},\textbf{1},0)(\textbf{1},\textbf{1},0)\leftrightarrow
N^{}_{Ri}~~(i=1,2,3)\,,
\end{eqnarray}
and an $[SU(2)^{}_L\times SU(2)'^{}_L]$-bidoublet scalar
\cite{flv1992,gu2012-2},
\begin{eqnarray}
\label{bhiggs}
\Sigma(\textbf{1},\textbf{2},+1)(\textbf{1},\textbf{2},+1)=\left[\begin{array}{ll}\sigma^{0}_{}&\sigma^{(-,0)}_{}\\
[1mm]
\sigma^{(0,-)}_{}&\sigma^{(-,-)}_{}\end{array}\right]\leftrightarrow
\Sigma\,,
\end{eqnarray}
where the first and second parentheses are the quantum numbers under
the $G$ and $G'$ gauge groups, respectively. Besides the gauge
symmetries and the mirror discrete symmetry, we impose a
$U(1)^{}_G\times U(1)'^{}_G$ global symmetry under which only the
$SU(2)$ doublets and the $SU(2)^{}_L\times SU(2)'^{}_L$ bidoublet
are nontrivial: $(1,0)$ for the $SU(2)^{}_L$ doublets, $(0,1)$ for
the $SU(2)'^{}_L$ doublets and $(1,1)$ for the $SU(2)^{}_L\times
SU(2)'^{}_L$ doublet.

\subsection{Interactions}

For simplicity, we only write down the following terms of the full
Lagrangian,
\begin{eqnarray}
\label{lagrangian} \mathcal{L}&\supset& -y_d^{}(\bar{q}^{}_{L}
\phi^{}_d d^{}_{R}+\bar{q}'^{}_{L} \phi'^{}_d d'^{}_{R})
-y_u^{}(\bar{q}^{}_{L} \phi^{}_u u^{}_{R}+\bar{q}'^{}_{L} \phi'^{}_u
u'^{}_{R})\nonumber\\
&& -y_e^{}(\bar{l}^{}_{L} \phi^{}_d e^{}_{R}+\bar{l}'^{}_{L}
\phi'^{}_d e'^{}_{R})-h(\bar{l}^{}_L\phi^{}_u N^{}_R+
\bar{l}'^{}_L\phi'^{}_u
N^{}_R)\nonumber\\
&&-\frac{1}{2}M_N^{}\bar{N}^c_R N^{}_R-f\bar{l}^{}_L \Sigma l'^{c}_L
- \rho \phi^\dagger_{u} \Sigma \phi'^\ast_{u}+\textrm{H.c.}\nonumber\\
&&-M_\Sigma^2
\textrm{Tr}(\Sigma^\dagger_{}\Sigma)-\frac{\epsilon}{2}B^{}_{\mu\nu}B'^{\mu\nu}_{}\,,
\end{eqnarray}
where $B^{}_{\mu\nu}$ and $B'^{}_{\mu\nu}$ are the $U(1)^{}_Y$ and
$U(1)'^{}_Y$ field strength tensors. Note the other gauge-invariant
trilinear couplings have been forbidden by the $U(1)^{}_G\times
U(1)'^{}_G$ global symmetry.

As a result of the mirror symmetry, the Yukawa couplings of the
$[SU(2)^{}_L\times SU(2)'^{}_L]$-bidoublet scalar to the
$[SU(2)]$-doublet leptons should have a symmetric structure,
\begin{eqnarray}
f=f^T_{}\,.
\end{eqnarray}
Furthermore, the mass term of the gauge-singlet fermions and the
trilinear coupling of the $[SU(2)^{}_L\times SU(2)'^{}_L]$-bidoublet
scalar to the $[SU(2)]$-doublet scalars softly break both of the
ordinary and dark lepton numbers. Without loss of generality and for
convenience, we will choose the base where the gauge-singlet
fermions have a diagonal and real mass matrix,
\begin{eqnarray}
M^{}_N=\textrm{diag}\{M_{N_1^{}}^{}, M_{N_2^{}}^{},
M_{N_3^{}}^{}\}\,,
\end{eqnarray}
to define three Majorana fermions,
\begin{eqnarray}
N_i^{}=N_{Ri}^{}+N_{Ri}^c\,.
\end{eqnarray}
Meanwhile, the $[SU(2)^{}_L\times SU(2)'^{}_L]$-bidoublet scalar can
have a real cubic coupling to the $[SU(2)^{}_L]$-doublet scalars by
a proper phase rotation, i.e.
\begin{eqnarray}
\rho=|\rho|\,.
\end{eqnarray}

\subsection{Vacuum expectation values}

We allow the discrete mirror symmetry and the global
$U(1)^{}_G\times U(1)'^{}_G$ symmetry to be softly broken by the
quadratic terms in the full scalar potential, which is not shown for
simplicity. So, the mirror Higgs scalars can develop the vacuum
expectation values (VEVs) different from those of the ordinary Higgs
scalars. In particular, the charged components of the mirror Higgs
scalars can have the nonzero VEVs \cite{gk2007,bcrwy2009}. In this
case, the symmetry breaking pattern should be
\begin{eqnarray}
SU(3)^{}_c\times SU(2)^{}_L\times U(1)^{}_Y&\longrightarrow&
SU(3)^{}_c\times U(1)^{}_{em}\,,\nonumber\\
SU(3)'^{}_c\times SU(2)'^{}_L\times U(1)'^{}_Y&\longrightarrow&
SU(3)'^{}_c\times U(1)'^{}_{em}\nonumber\\
&\longrightarrow& SU(3)'^{}_c\,.
\end{eqnarray}
Accordingly, the mirror photon can become massive although the
ordinary photon keeps massless.

The VEVs of the $[SU(2)^{}_L]$-doublet Higgs scalars $\phi^{}_u$ and
$\phi^{}_d$ should be fixed by \cite{gk2007}
\begin{eqnarray}
\langle\phi^{}_{u}\rangle=\left[\begin{array}{c}\langle\phi^0_{u}\rangle\\
[2mm] 0\end{array}\right]\,,~
\langle\phi^{}_{d}\rangle=\left[\begin{array}{c}0\\
[2mm]\langle\phi^0_{d}\rangle\end{array}\right]~\textrm{with}\quad\quad\quad\quad\quad&&\nonumber\\
\langle\phi\rangle^{2}_{}=\sqrt{\langle\phi^{}_{u}\rangle^{2}_{}
+\langle\phi^{}_{d}\rangle^{2}_{}}\simeq 174\,\textrm{GeV}\,,&&
\end{eqnarray}
while the VEVs of the $[SU(2)'^{}_L]$-doublet Higgs scalars
$\phi'^{}_u$ and $\phi'^{}_d$ can be described by \cite{gk2007}
\begin{eqnarray}
\label{chargedvev}
&&\!\!\!\!\!\!\!\!\!\langle\phi'^{}_{u}\rangle=\left[\begin{array}{c}\langle\phi'^{0}_{u}\rangle\\
[2mm] 0\end{array}\right]\,,~
\langle\phi'^{}_{d}\rangle=\left[\begin{array}{c}\langle\phi'^{+}_{d}\rangle\\
[2mm]\langle\phi'^{0}_{d}\rangle\end{array}\right]\,.\nonumber\\
&&
\end{eqnarray}
The $[SU(2)^{}_L\times SU(2)'^{}_L]$-bidoublet Higgs scalar $\Sigma$
can pick up the VEV through its cubic coupling to the
$[SU(2)]$-doublet Higgs scalars $\phi^{}_u$ and $\phi'^{}_u$,
\begin{eqnarray}
\langle\Sigma\rangle=\left[\begin{array}{cc}\langle\sigma^0_{}\rangle&0\\
[2mm]
0&0\end{array}\right]~~\textrm{with}~~\langle\sigma^0_{}\rangle\simeq
-\frac{\rho \langle\phi^0_{u}\rangle
\langle\phi'^0_{u}\rangle}{M_{\Sigma}^2}\,.
\end{eqnarray}
Clearly, the VEV $\langle\Sigma\rangle$ can be much smaller than the
VEVs $\langle\phi^0_{u}\rangle$ and $\langle\phi'^0_{u}\rangle$ in
the seesaw scenario, i.e.
\begin{eqnarray}
\langle\Sigma\rangle\ll
\langle\phi^0_{u}\rangle\,,~\langle\phi'^0_{u}\rangle~~\textrm{for}~~\langle\phi^0_u\rangle\,,~\langle\phi'^0_u\rangle
\ll \rho \lesssim M_{\Sigma}^{}\,.
\end{eqnarray}

\subsection{Mirror photon}

We can make a non-unitary transformation \cite{fh1991},
\begin{eqnarray}
\tilde{B}^{}_\mu=B^{}_\mu+\epsilon B'^{}_\mu\,,
~~\tilde{B}'^{}_\mu=\sqrt{1-\epsilon^2}B'^{}_\mu\,,
\end{eqnarray}
to remove the $U(1)^{}_Y\times U(1)'^{}_Y$ kinetic mixing and then
define the orthogonal fields,
\begin{eqnarray}
\begin{array}{l}
A^{}_\mu =W^3_\mu \sin\theta_W^{} +\tilde{B}^{}_\mu \cos\theta_W^{}\,,\\
[2mm] Z^{}_\mu =W^3_\mu \cos\theta^{}_W -\sin\tilde{B}^{}_\mu
\theta^{}_W\,,\\
[2mm]A'^{}_\mu =W'^3_\mu \sin\theta^{}_{W} +\tilde{B}'^{}_\mu
\cos\theta^{}_{W}\,,\\
[2mm] Z'^{}_\mu =W'^3_\mu \cos\theta^{}_{W} -\tilde{B}'^{}_\mu
\sin\theta^{}_{W}\,.\end{array}
\end{eqnarray}
Here $\theta_W^{}$ with $\sin^2_{}\theta_W^{}\simeq 0.231$ is the
Weinberg angle while $W^{3}_{\mu}$ and $W'^{3}_{\mu}$ are the
$SU(2)^{}_L$ and $SU(2)'^{}_L$ gauge fields. In the above orthogonal
base, the field $A^{}_\mu$ is exactly massless and is the physical
mass-eigenstate field, the ordinary photon, according to the
unbroken electromagnetic symmetry $U(1)^{}_{em}$ in the ordinary
sector, while the others $Z^{}_\mu$, $Z'^{}_\mu$ and $A'^{}_\mu$
will mix together. The mirror electromagnetic symmetry
$U(1)'^{}_{em}$ is broken by the charged VEV
$\langle\phi^{+}_d\rangle$ given in Eq. (\ref{chargedvev}).
Consequently, the $W'^{\pm}_{}$ boson will also mix with the $Z'$
boson and the mirror photon $A'^{}_\mu$, which is massive now.

The mirror photon can couple to the ordinary fermions besides the
mirror fermions,
\begin{eqnarray}
\mathcal{L}&\supset& eA'^{}_\mu\left\{\frac{\epsilon
}{4}\left[\bar{e}\gamma^\mu_{}
\left(3+\gamma_5^{}\right)e+\bar{\nu}\gamma^\mu_{}(1-\gamma_5^{})
\nu\right.\right.\nonumber\\
&&\left.+\bar{d}\gamma^\mu_{}
\left(\frac{1}{3}+\gamma_5^{}\right)d-\bar{u}\gamma^\mu_{}
\left(\frac{5}{3}+\gamma_5^{}\right)u\right]\nonumber\\
&&\left.+\left(-\frac{1}{3}\bar{d}'\gamma^\mu_{}d'+\frac{2}{3}\bar{u}'\gamma^\mu_{}u'
-\bar{e}'\gamma^\mu_{}e'\right)\right\}~~\textrm{for}~~\epsilon\ll
1\,.\nonumber\\
&&
\end{eqnarray}
In the case with
$\langle\phi'^{+}_d\rangle=\mathcal{O}(100\,\textrm{MeV})$, the
mirror photon can have a mass
\begin{eqnarray}
m_{A'}^{}\simeq
\sqrt{8\pi\alpha}\langle\phi'^{+}_d\rangle=\mathcal{O}(100\,\textrm{MeV})\,,
\end{eqnarray}
and its decay width will not be smaller than
\begin{eqnarray}
\label{mphoton} \Gamma_{A'}^{}&=&\Gamma_{A'\rightarrow
\nu\bar{\nu}}^{}+\Gamma_{A'\rightarrow
e\bar{e}}^{}+\Gamma_{A'\rightarrow
u\bar{u}}^{}+\Gamma_{A'\rightarrow
d\bar{d}}^{}\nonumber\\
&\simeq &\frac{5}{9}\epsilon^2_{}\alpha m_{A'}^{}\,.
\end{eqnarray}
Here $\alpha=e^2_{}/(4\pi)\simeq 1/137$ \cite{nakamura2010} is the
fine structure constant.

\section{Seesaw for active and sterile neutrino masses}

From Eq. (\ref{lagrangian}), the ordinary active neutrinos, the
mirror sterile neutrinos and the gauge-singlet fermions will have
the mass matrix as below,
\begin{eqnarray}
\mathcal{L}\!\!&\supset&\!\!-\frac{1}{2}\left[\begin{array}{ccc}\bar{\nu}^{}_L&\bar{\nu}'^{}_L
&\bar{N}^{c}_R\end{array}\right]\!\!
\left[\begin{array}{ccc}0& f\langle\Sigma\rangle& h \langle\phi^{}_u\rangle\\
[2mm]f^T_{}\langle\Sigma\rangle& 0& h  \langle\phi'^{}_u\rangle\\
[2mm]h^T_{}  \langle\phi^{}_u\rangle & h^T_{}
\langle\phi'^{}_u\rangle & M^{}_N\end{array}\right]\!\!
\left[\begin{array}{c}\nu^{c}_L\\
[2mm] \nu'^{c}_L\\
[2mm] N^{}_R\end{array}\right]\nonumber\\
&&+\textrm{H.c.}\,,
\end{eqnarray}
after the ordinary and mirror electroweak symmetry breaking.

\subsection{Active and sterile neutrino masses and mixing}

As long as the gauge-singlet fermions are heavy enough, i.e.
\begin{eqnarray}
M^{}_N \gg h
\langle\phi^{}_u\rangle\,,~h\langle\phi'^{}_u\rangle\,,~
f\langle\Sigma\rangle\,,
\end{eqnarray}
we can make use of the seesaw mechanism to get the mass matrix of
the active and sterile neutrinos,
\begin{widetext}
\begin{subequations}
\begin{eqnarray}
\label{asmatrixf}
\mathcal{L}&\supset&-\frac{1}{2}\left[\begin{array}{cc}\bar{\nu}^{}_L&\bar{\nu}'^{}_L\end{array}\right]
\left[\begin{array}{cc}-h\frac{\langle\phi^{}_u\rangle^2_{}}{M_N^{}}h^T_{}
& f\langle\Sigma\rangle-
h\frac{\langle\phi^{}_u\rangle\langle\phi'^{}_u\rangle}{M_N^{}}h^T_{}\\
[4mm]f\langle\Sigma\rangle-
h\frac{\langle\phi^{}_u\rangle\langle^{}_u\phi'\rangle}{M_N^{}}h^T_{}&
-h\frac{\langle\phi'^{}_u\rangle^2_{}}{M_N^{}}h^T_{}
\end{array}\right]
\left[\begin{array}{c}\nu^{c}_L\\
[4mm] \nu'^{c}_L\end{array}\right]+\textrm{H.c.}\,,\\
[2mm]\label{asmatrixm}
&=&-\frac{1}{2}\left[\begin{array}{cc}\bar{\nu}^{}_L&\bar{\nu}'^{}_L\end{array}\right]
\left[\begin{array}{cc}U_{\nu}^{}& U_{\nu\nu'}^{}\\
[2mm] U_{\nu'\nu}^{}&U_{\nu'}^{}
\end{array}\right]\left[\begin{array}{cc}\hat{m}_\nu^{}&0\\
[2mm] 0&\hat{m}_{\nu'}^{}
\end{array}\right]\left[\begin{array}{cc}U_{\nu}^{T}& U_{\nu'\nu}^{T}\\
[2mm] U_{\nu\nu'}^{T}&U_{\nu'}^{T}
\end{array}\right]
\left[\begin{array}{c}\nu^{c}_L\\
[2mm] \nu'^{c}_L\end{array}\right]+\textrm{H.c.}\,.
\end{eqnarray}
\end{subequations}
\end{widetext}
Here the mass eigenvalues have been introduced,
\begin{eqnarray}
\hat{m}_{\nu}^{}&=&\textrm{diag}\{m_{\nu_1^{}}^{},m_{\nu_2^{}}^{},m_{\nu_3^{}}^{}\}\,,\nonumber\\
\hat{m}_{\nu'}^{}&=&\textrm{diag}\{m_{\nu'^{}_1}^{},m_{\nu'^{}_2}^{},m_{\nu'^{}_3}^{}\}\,.
\end{eqnarray}
If the entries in the mass matrix (\ref{asmatrixf}) have the
following hierarchy,
\begin{eqnarray}
\label{swcondition}
-h\frac{\langle\phi'^{}_u\rangle^2_{}}{M_N^{}}h^T_{}\gg
f\langle\Sigma\rangle-h\frac{\langle\phi^{}_u\rangle\langle\phi'^{}_u\rangle}{M_N^{}}h^T_{}\,,
~-h\frac{\langle\phi^{}_u\rangle^2_{}}{M_N^{}}h^T_{}\,,
\end{eqnarray}
the sterile neutrino masses should have a form of the canonical
seesaw,
\begin{eqnarray}
\label{sterile} \mathcal{L}\supset-\bar{\nu}'^{}_L
m_{\nu'}^{}\nu'^c_L+\textrm{H.c.}~~\textrm{with}~~m_{\nu'}^{}=-h\frac{\langle\phi'^{}_u\rangle^2_{}}{M_N^{}}h^T_{}\,,
\end{eqnarray}
while the active neutrino masses should have a form of the double
and linear seesaw,
\begin{eqnarray}
\label{active}
\mathcal{L}&\supset&f\frac{\langle\Sigma\rangle^2_{}}{h\frac{\langle\phi'^{}_u\rangle^2_{}}{M_N^{}}h^T_{}}f
-2f\frac{\langle\phi^{}_u\rangle\langle\Sigma\rangle}{\langle\phi'^{}_u\rangle}\,.
\end{eqnarray}
Under the seesaw condition (\ref{swcondition}), the active mixing
matrix $U_{\nu}^{}$ and the sterile mixing matrix $U_{\nu'}^{}$ can
approximate to the Pontecorvo-Maki-Nakagawa-Sakata \cite{mns1962}
(PMNS) matrices,
\begin{eqnarray}
\label{pmns} &&U_{\nu}^{\dagger}U_{\nu}^{}=
U_{\nu}^{}U_{\nu}^{\dagger}= 1\,,~~U_{\nu'}^{\dagger}U_{\nu'}^{}=
U_{\nu'}^{}U_{\nu'}^{\dagger}= 1\,,
\end{eqnarray}
while the active-sterile mixing matrices $U_{\nu\nu'}$ and
$U_{\nu'\nu}$ can be simplified by
\begin{eqnarray}
\label{asmixing} U_{\nu\nu'}^{}= f
U^\ast_{\nu'}\frac{\langle\Sigma\rangle}{\hat{m}_{\nu'}^{}}
+\frac{\langle\phi^{}_u\rangle}{\langle\phi'^{}_u\rangle}U_{\nu'}^{\dagger}\,,~~U_{\nu'\nu}^{}=
-U_{\nu'}^{\dagger}U_{\nu\nu'}^{\dagger}U_{\nu}^{\dagger}\,.
\end{eqnarray}

\subsection{Sterile neutrino decays}

Due to their mixing with the ordinary neutrinos, the sterile
neutrinos can decay into the ordinary fermions \cite{bhl2009},
\begin{subequations}
\label{mndecay}
\begin{eqnarray}
\Gamma_{\nu'^{}_i\rightarrow  \nu\nu\nu}^{}&=&\frac{ G_F^2
m_{\nu'^{}_i}^5}{96\pi^3}(U_{\nu\nu'}^{\dagger}U_{\nu\nu'}^{})_{ii}\,,\\
\Gamma_{\nu'^{}_i\rightarrow  \nu e^{+}_{}e^{-}_{}}^{}&=&\frac{5
G_F^2
m_{\nu'^{}_i}^5}{768\pi^3}(U_{\nu\nu'}^{\dagger}U_{\nu\nu'}^{})_{ii}\,,\\
\Gamma_{\nu'^{}_i\rightarrow  \nu u\bar{u}}^{}&=&\frac{G_F^2
m_{\nu'^{}_i}^5}{32\pi^3}(1-\frac{8}{3}s^2_W
+\frac{32}{9}s^4_W)(U_{\nu\nu'}^{\dagger}U_{\nu\nu'}^{})_{ii}\,,\nonumber\\
&& \\
\Gamma_{\nu'^{}_i\rightarrow  \nu d\bar{d}}^{}&=&\frac{G_F^2
m_{\nu'^{}_i}^5}{32\pi^3}(1-\frac{4}{3}s^2_W
+\frac{8}{9}s^4_W)(U_{\nu\nu'}^{\dagger}U_{\nu\nu'}^{})_{ii}\,,\nonumber\\
&& \\
\Gamma_{\nu'^{}_i\rightarrow  e^{-}_{}u\bar{d}}^{}&=&
\Gamma_{\nu'^{}_i\rightarrow
e^{+}_{}d\bar{u}}^{}=\frac{|V_{ud}^{}|^2_{}G_F^2
m_{\nu'^{}_i}^5}{32\pi^3}(U_{\nu\nu'}^{\dagger}U_{\nu\nu'}^{})_{ii}\,,\nonumber\\
&&
\end{eqnarray}
\end{subequations}
if the kinematics is allowed. Here $G_F^{}=1.16637 \times
10^{-5}_{}\,\textrm{GeV}^{-2}_{}$ is the Fermi constant,
$s_W^2=\sin^2_{}\theta_W^{}$ is the Weinberg angle, while
$V_{ud}^{}\simeq 0.97419$ is an element of the
Cabibbo-Kobayashi-Maskawa matrix \cite{nakamura2010}.

\section{Leptogenesis for ordinary and mirror baryon asymmetries}

\begin{figure*}
\vspace{6.5cm} \epsfig{file=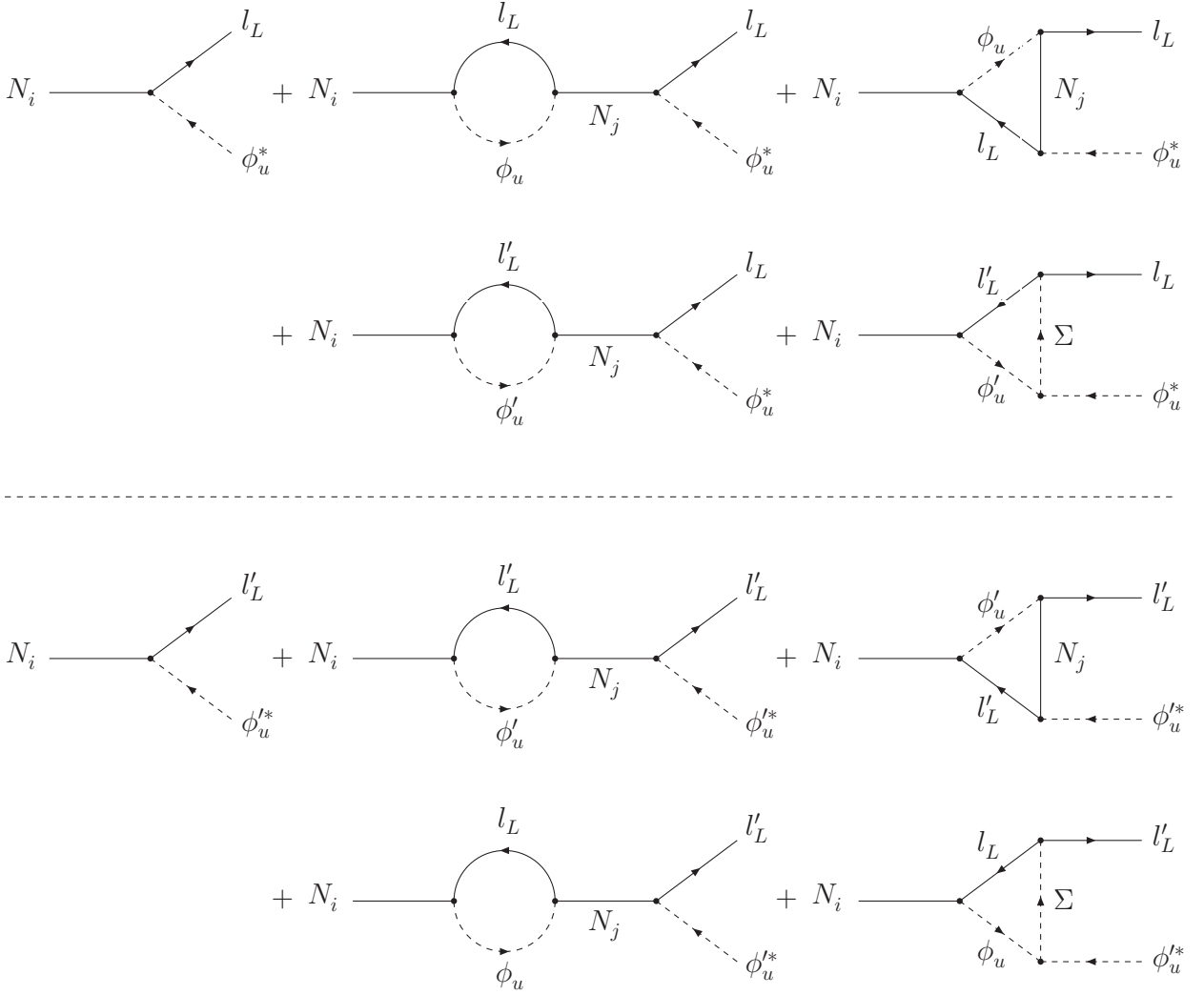, bbllx=5.5cm, bblly=6.0cm,
bburx=15.5cm, bbury=16cm, width=8cm, height=8cm, angle=0, clip=0}
\vspace{0cm} \caption{\label{Ndecay} The heavy Majorana fermions
$(N_i^{}=N^{}_{Ri}+N^{c}_{Ri}~\textrm{with}~N^{}_{Ri}~\textrm{being~three~gauge~singlets})$
decays into the $[SU(2)^{}_L]$-doublet leptons and Higgs scalar
$(l^{}_L,\phi^{}_u)$ as well as into the $[SU(2)'^{}_L]$-doublet
dark leptons and Higgs scalar $(l'^{}_L,\phi'^{}_u)$. Here $\Sigma$
is a heavy $[SU(2)^{}_L\times SU(2)'^{}_L]$-bidoublet scalar. }
\end{figure*}

If CP is not conserved, the decays of the heavy Majorana fermions
$N^{}_i$ can simultaneously generate two types of lepton
asymmetries: one is stored in the $[SU(2)^{}_L]$-doublet leptons
$l^{}_L$, i.e.
\begin{eqnarray}
\eta^{}_L=\frac{n_{l_L^{}}^{}}{s}\,,
\end{eqnarray}
and the other is stored in the mirror leptons $l'^{}_L$, i.e.
\begin{eqnarray}
\eta'^{}_L=\frac{n_{l'^{}_L}^{}}{s}\,.
\end{eqnarray}
Here $n_{l^{}_L}^{}$ and $n_{l'^{}_L}^{}$ are the number densities
while $s$ is the entropy density. The relevant diagrams are shown in
Fig. \ref{Ndecay}.

The $SU(2)^{}_L$ sphaleron processes \cite{krs1985} then will
partially transfer the ordinary lepton asymmetry to an ordinary
baryon asymmetry,
\begin{eqnarray}
\eta^{}_B=-\frac{28}{79}\eta^{}_L\,.
\end{eqnarray}
Similarly, the mirror lepton asymmetry will be partially converted
to a mirror baryon asymmetry,
\begin{eqnarray}
\eta'^{}_B=-\frac{28}{79}\eta'^{}_L\,,
\end{eqnarray}
through the $SU(2)'^{}_L$ sphaleron processes \cite{krs1985}. Due to
the Yukawa couplings in Eq. (\ref{lagrangian}), the ordinary lepton
asymmetry and then the ordinary baryon asymmetry should equal to the
mirror ones, i.e.
\begin{eqnarray}
\label{obmb} \eta^{}_L=\eta'^{}_L\propto
\varepsilon_{N_i^{}}^{}\Rightarrow \eta^{}_B=\eta'^{}_B\propto
\varepsilon_{N_i^{}}^{}\,.
\end{eqnarray}
Here $\varepsilon_{N_i^{}}^{}$ is the CP asymmetry in the decays of
the heavy Majorana fermions $N^{}_i$.

\subsection{CP violation in decays}

The total decay width in the decays of the heavy Majorana fermions
$N^{}_i$ can be calculated at tree level,
\begin{eqnarray}
\Gamma_{N_i^{}}^{}&=&\Gamma_{N_i^{}\rightarrow
l_L^{}\phi^\ast_u}^{}+\Gamma_{N_i^{}\rightarrow
l'^{c}_L\phi'^{}_u}^{}+\Gamma_{N_i^{}\rightarrow
l'^{}_L\phi'^\ast_{u}}^{}+\Gamma_{N_i^{}\rightarrow
l^c_L\phi^{}_u}^{}\nonumber\\
&=&\frac{1}{4\pi}(h^\dagger_{}h)_{ii}^{}M_{N_i^{}}^{}\,.
\end{eqnarray}
We then can compute the CP asymmetry $\varepsilon_{N_i^{}}^{}$
appeared in Eq. (\ref{obmb}) at one-loop level,
\begin{widetext}
\begin{eqnarray}
\varepsilon_{N_i^{}}^{}&=&\frac{\Gamma_{N_i^{}\rightarrow
l_L^{}\phi^\ast_{u}}^{}-\Gamma_{N_i^{}\rightarrow
l_L^{c}\phi^{}_u}^{}}{\Gamma_{N_i^{}}^{}}=\frac{\Gamma_{N_i^{}\rightarrow
l'^{}_{L}\phi'^\ast_{u}}^{}-\Gamma_{N_i^{}\rightarrow
l'^{c}_{L}\phi'^{}_{u}}^{}}{\Gamma_{N_i^{}}^{}}\nonumber\\
&=&\frac{1}{16\pi}\frac{1}{(h_{}^\dagger
h)_{ii}^{}}\left\{\sum_{j\neq i}^{}\textrm{Im}\left[(h_{}^\dagger
h)_{ij}^{2}\right]\left[2S\left(\frac{M_{N_j^{}}^{2}}{M_{N_i^{}}^{2}}\right)
+V\left(\frac{M_{N_j^{}}^{2}}{M_{N_i^{}}^{2}}\right)\right]+\textrm{Im}\left[(h_{}^\dagger
fh_{}^{\ast})_{ii}^{}\right]
\tilde{V}\left(\frac{M_{\Sigma}^{2}}{M_{N_i^{}}^{2}},\frac{\rho^2_{}}{M_{\Sigma}^{2}}\right)\right\}\,,
\end{eqnarray}
\end{widetext}
with $S(x)$ being the self-energy correction, $V(x)$ and
$\tilde{V}(x,y)$ being the vertex corrections,
\begin{subequations}
\begin{eqnarray}
S(x)&=&\frac{\sqrt{x}}{1-x}\,,\\
V(x)&=&\sqrt{x}\left[1-(1+x)\ln\left(\frac{1+x}{x}\right)\right]\,,\\
\tilde{V}(x,y)&=&2\sqrt{xy}\left[-1+x\ln\left(\frac{1+x}{x}\right)\right]\,.
\end{eqnarray}
\end{subequations}
In the limit $M_{N_i^{}}^2\ll M_{N_j^{}}^2, M_{\Sigma}^2$, the CP
asymmetry $\varepsilon_{N_i^{}}^{}$ can be simplified as
\begin{eqnarray}
\label{scpa}
\varepsilon_{N_i^{}}^{}\simeq\frac{5}{32\pi}\frac{\textrm{Im}\left[h^\dagger_{}\left(m_{\nu'}^{}
-\frac{\langle\phi'^{}_u\rangle}{5\langle\phi^{}_u\rangle}m_{\nu}^{\textrm{Linear}}\right)h^\ast_{}\right]_{ii}^{}
M_{N_i^{}}^{}}
{(h_{}^\dagger h)_{ii}^{}\langle\phi'^{}_u\rangle^2_{}}\,.
\end{eqnarray}

As we will show later the dark matter relic density and the BNN
constraint enforce
\begin{eqnarray}
\label{cl} m_{\nu'}^{\textrm{max}}\gg
-\frac{\langle\phi'^{}_u\rangle}{5\langle\phi^{}_u\rangle}m_{\nu}^{\textrm{Linear}}\,,
\end{eqnarray}
with $m_{\nu'}^{\textrm{max}}$ being the maximal mass eigenvalue of
the mirror neutrinos. So, the simplified CP asymmetry (\ref{scpa})
can have an upper bound,
\begin{eqnarray}
\label{cpamax}
\left|\varepsilon_{N_i^{}}^{}\right|<\varepsilon_{N_i^{}}^{\textrm{max}}=
\frac{5}{32\pi}\frac{ M_{N_i^{}}^{} m_{\nu'}^{\textrm{max}}}
{\langle\phi'^{}_u\rangle^2_{}}\,,
\end{eqnarray}
which is similar to the Davidson-Ibarra bound \cite{di2002} in the
canonical seesaw scenario.

\subsection{Scattering processes}

The Majorana fermions $N_i^{}$ and the Higgs bidoublet $\Sigma$ can
mediate some lepton-number-violating scattering processes such as
$l^{}_L\phi^\ast_{u}\rightarrow l^c_L \phi^{}_u$,
$l'^{}_L\phi'^\ast_{u}\rightarrow l'^c_L \phi'^{}_u$,
$l^{}_L\phi^\ast_{u}\rightarrow l'^c_L \phi'^{}_u$,
$l'^{}_L\phi'^\ast_{u}\rightarrow l^c_L \phi^{}_u$ and so on. These
scattering processes will not be kept in equilibrium below the
temperature $T_D^{}$ given by \cite{kt1990}
\begin{eqnarray}
\left[\Gamma_S^{}\simeq H(T)\right]\left|_{T=T_D^{}}^{}\right.\,.
\end{eqnarray}
Here $\Gamma_S^{}$ is the interaction rate of the scattering
processes, while
\begin{eqnarray}
H(T)=\left(\frac{8\pi^{3}_{}g_{\ast}^{}}{90}\right)^{\frac{1}{2}}_{}
\frac{T^{2}_{}}{M_{\textrm{Pl}}^{}}\,,
\end{eqnarray}
is the Hubble constant with $M_{\textrm{Pl}}^{}\simeq 1.22\times
10^{19}_{}\,\textrm{GeV}$ being the Planck mass and
$g_\ast^{}=2\times(106.75+2)=217.5$ being the relativistic degrees
of freedom. Below the masses of the mediators $N_i^{}$ and $\Sigma$,
the interaction rate can be given by \cite{fy1990}
\begin{eqnarray}
\Gamma_S^{}&=&\frac{2}{\pi^3_{}}\frac{T^3_{}}{\langle\phi'^{}_u\rangle^4_{}}\textrm{Tr}\left[m_{\nu'}^\dagger
m_{\nu'}^{}+\left(m_{\nu'}^{}
+\frac{\langle\phi'^{}_u\rangle}{2\langle\phi^{}_u\rangle}m_{\nu}^{\textrm{Linear}}\right)^\dagger_{}\right.\nonumber\\
&&\left.\times \left(m_{\nu'}^{}
+\frac{\langle\phi'^{}_u\rangle}{2\langle\phi^{}_u\rangle}m_{\nu}^{\textrm{Linear}}\right)\right]~~\textrm{for}~~T\ll
M_{N}^{}\,,M_{\Sigma}^{}\,,\nonumber\\
&&
\end{eqnarray}
which tends to
\begin{eqnarray}
\label{srate}
&&\Gamma_S^{}\simeq\frac{4}{\pi^3_{}}\frac{T^3_{}}{\langle\phi'^{}_u\rangle^4_{}}\textrm{Tr}\left(m_{\nu'}^\dagger
m_{\nu'}^{}\right)=\frac{4}{\pi^3_{}}\frac{T^3_{}}{\langle\phi'^{}_u\rangle^4_{}}\sum_{i}^{}m_{\nu'^{}_{i}}^2\nonumber\\
&&\Rightarrow T_D^{}\simeq
\frac{\pi^{\frac{9}{2}}_{}g_\ast^{\frac{1}{2}}}{6\sqrt{5}}\frac{\langle\phi'^{}_u\rangle^4_{}}
{M_{\textrm{Pl}}^{}\sum_i^{}m_{\nu'^{}_i}^2}\,,
\end{eqnarray}
for the constraint (\ref{cl}). Alternatively, the scattering
processes can decouple at a  temperature above or near the
mediator's mass if the interactions are weak enough to satisfy
\cite{kt1990}
\begin{eqnarray}
\label{equilibrium} K_{N_i^{}}^{}=\frac{\Gamma_{N_i^{}}^{}}{2
H(T)}\left|_{T=M_{N_i^{}}^{}}^{}\right.\ll
1\,,&&\!\!\!\!\!\!K_{\Sigma}^{}=\frac{\Gamma_{\Sigma}^{}}{2
H(T)}\left|_{T={M_\Sigma^{}}^{}}^{}\right.\ll 1\,.\nonumber\\
&&
\end{eqnarray}

\subsection{Final baryon asymmetries}

In the case the lightest Majorana fermion $N_1^{}$ has a mass
smaller than the decouple temperature of the scattering processes
mediated by the other Majorana fermions $N_{2,3}^{}$ and the Higgs
bidoublet $\Sigma$, i.e.
\begin{eqnarray}
M_{N_1^{}}^{}< T_D^{}\,,
\end{eqnarray}
the final baryon asymmetries can be approximately solved by
\cite{kt1990}
\begin{eqnarray}
\label{basymmetry} \eta_B^{}= \eta'^{}_B&\simeq&
-\frac{28}{79}\times \frac{\varepsilon_{N_1^{}}^{}}{g_\ast^{}}\times
\kappa\nonumber\\
&=&0.888\times 10^{-10}_{}
\left(\frac{\varepsilon_{N_1^{}}^{}}{-5.45\times
10^{-8}_{}}\right)\left(\frac{\kappa}{1}\right)~~\textrm{with}\nonumber\\
&&\kappa\simeq\left\{\begin{array}{cl}1 &
K_{N_1^{}}^{}\ll 1\,,\\
[3mm] \frac{0.3}{K_{N_1^{}}^{}(\ln
K_{N_1^{}}^{})^{0.6}_{}}&K_{N_1^{}}^{}\gtrsim 1\,.\end{array}\right.
\end{eqnarray}

\section{Implications and constraints}

Before giving the concrete parameter choice, we shall demonstrate
some general implications and constraints on the model.

\subsection{Dark matter mass}

From the Yukawa couplings in Eq. (\ref{lagrangian}), we can easily
read the relation between the ordinary and mirror charged fermion
masses,
\begin{eqnarray}
\frac{\langle\phi'^{}_d\rangle}{\langle\phi^{}_d\rangle}&=&\frac{m_{d'}^{}}{m_{d}^{}}=\frac{m_{s'}^{}}{m_{s}^{}}
=\frac{m_{b'}^{}}{m_{b}^{}}=\frac{m_{e'}^{}}{m_{e}^{}}=\frac{m_{\mu'}^{}}{m_{\mu}^{}}
=\frac{m_{\tau'}^{}}{m_{\tau}^{}}\,,\nonumber\\
\frac{\langle\phi'^{}_u\rangle}{\langle\phi^{}_u\rangle}&=&\frac{m_{u'}^{}}{m_{u}^{}}=\frac{m_{c'}^{}}{m_{c}^{}}
=\frac{m_{t'}^{}}{m_{t}^{}}\,.
\end{eqnarray}
As a result of the mirror symmetry, the ordinary and mirror strong
coupling constants should become equal at sufficiently high scales.
Therefore, the beta functions of the ordinary and mirror QCD can
govern the dependence of the mirror hadronic scale on the ordinary
one \cite{acmy2009},
\begin{eqnarray}
\label{mqcd} \Lambda^{}_{\textrm{QCD}'}&=&
\left(\frac{\langle\phi'^{}_u\rangle}{\langle\phi^{}_u\rangle}\right)^{\frac{4}{11}}_{}
\left(\!\frac{\tan\beta}{\tan\beta'}\right)^{\frac{2}{11}}_{}
(m_u^{}m_d^{}m_s^{})^{\frac{2}{33}}_{}\Lambda^{\frac{9}{11}}_{\textrm{QCD}}\nonumber\\
&&~~\textrm{for}~~\Lambda^{}_{\textrm{QCD}'}<m_{u'}^{}\,,m_{u'}^{}\,,
\end{eqnarray}
where we have defined
\begin{eqnarray}
\tan\beta=\frac{\langle\phi^{}_u\rangle}{\langle\phi^{}_d\rangle}\,,
~~\tan\beta'&=&\frac{\langle\phi'^{}_u\rangle}{\langle\phi'^{}_d\rangle}\,.
\end{eqnarray}

In the ordinary sector, the current quark masses $m_u^{}$ and
$m_d^{}$ are much smaller than the hadronic scale
$\Lambda_{\textrm{QCD}}^{}$ so that they can only have a negligible
contribution to the nucleon masses,
\begin{eqnarray}
m_p^{}\simeq m_n^{}\simeq 940\,\textrm{MeV}=m_{N}^{}\,.
\end{eqnarray}
In addition, the $\Delta$ baryons and the neutron has a mass split
from the hyperfine interaction among the constituent quarks
\cite{acmy2009},
\begin{eqnarray}
m_{\Delta}^{}-m_{n}^{}\simeq 300\,\textrm{MeV} \propto
\frac{\Lambda_{\textrm{QCD}}^{3}}{m_q^2}\,,
\end{eqnarray}
with $m_q^{}\simeq 300\,\textrm{MeV}$ being the constituent quark
mass. In the mirror sector, the quark masses $m_{u'}^{}$ and
$m_{d'}^{}$ may be larger than the hadronic scale
$\Lambda^{}_{\textrm{QCD}'}$. The mirror proton and neutron masses
then can approximately equal to the sum of the mirror quark masses,
\begin{eqnarray}
\label{mpnmass} m_{p'}^{}=2m_{u'}^{}+m_{d'}^{}\,,~~
m_{n'}^{}=2m_{d'}^{}+ m_{u'}^{}\,,
\end{eqnarray}
which implies
\begin{subequations}
\begin{eqnarray}
\label{mprotn} m_{p'}^{}&<&
m_{n'}^{}~\textrm{for}~m_{u'}^{}< m_{d'}^{}\,,\quad\\
[2mm] \label{mneutron} m_{p'}^{}&>&
m_{n'}^{}~\textrm{for}~m_{u'}^{}>m_{d'}^{}\,.\quad
\end{eqnarray}
\end{subequations}
In the case (\ref{mneutron}), the mirror $\Delta'^{-}_{}$ baryon can
be lighter than the mirror neutron if the mirror hyperfine
interaction doesn't compensate the mass difference
$m_{u'}^{}-m_{d'}^{}$ \cite{acmy2009},
\begin{eqnarray}
\label{mbmass}
m_{\Delta'^{-}_{}}^{}=3m_{d'}^{}+(m_{\Delta}^{}-m_{n}^{})
\left(\frac{\Lambda_{\textrm{QCD}'}^{}}{\Lambda_{\textrm{QCD}}^{}}\right)^3_{}\!\!\!
\left(\frac{m_q^2}{m_{u'}^{}m_{d'}}\right).
\end{eqnarray}

As the lightest mirror baryon is expected to serve as the dark
matter particle, its mass should be
\begin{eqnarray}
\label{dmmass} m_{\textrm{DM}}^{}&\simeq &5\,m_N^{}\simeq
5\,\textrm{GeV}\,,
\end{eqnarray}
to explain the cosmological observations,
\begin{eqnarray}
\Omega_B^{}h^2_{}:\Omega_{\textrm{DM}}^{}h^2_{}&=&m_N^{}\eta_B^{}:m_{\textrm{DM}}^{}\eta'^{}_{B}
=m_N^{}:m_{\textrm{DM}}^{}\nonumber\\
&\simeq& 1:5\,.
\end{eqnarray}

\subsection{Dark matter direct detection}

In the presence of the $U(1)^{}_Y\times U(1)'^{}_Y$ kinetic mixing,
the mirror photon can mediate a scattering of the dark matter
particle off the ordinary nucleons. For example, the mirror proton
$p'$ or the mirror $\Delta'^{-}_{}$ baryon has a spin-independent
cross section,
\begin{eqnarray}
\sigma_{X N\rightarrow X N}^{}&\simeq& \epsilon^2_{}\frac{\pi\,
\alpha^2_{}\mu_r^2}{m_{A'}^4}
\left[\frac{3Z+(A-Z)}{A}\right]^2_{}\,,\nonumber\\
&\simeq &
10^{-41}_{}\,\textrm{cm}^2_{}\left(\frac{\epsilon}{1.5\times
10^{-7}_{}}\right)^2_{}\nonumber\\
&&\times\left(\frac{\mu_r^{}}{0.833\,\textrm{GeV}}\right)^2_{}
\left(\frac{100\,\textrm{MeV}}{m_{A'}^{}}\right)^4_{}\nonumber\\
&&\times \left[\frac{3Z+(A-Z)}{A}\right]^2_{}\,,
\end{eqnarray}
which can be close to the XENON10 limit \cite{angle2011}. Here $X$
denotes the mirror proton $p'$ or the mirror $\Delta'^{-}_{}$
baryon, $Z$ and $A-Z$ are the numbers of proton and neutron within
the target nucleus, while
\begin{eqnarray}
\mu_r^{}&=&\frac{m_{X}^{} m_N^{}}{m_{X}^{}+m_N^{}}\simeq
0.833\,\textrm{GeV}\nonumber\\
&&~~\textrm{for}~~m_{X}^{}\simeq 5\,m_N^{}\simeq 5\,\textrm{GeV}\,,
\end{eqnarray}
is the reduced mass. Alternatively, the mirror neutron $n'$ can
serve as the dark matter particle if it is the lightest mirror
baryon. The mirror neutron as the dark matter particle can have an
energy-dependent cross section. The detailed studies can be found in
\cite{acmy2009}.

\subsection{Constraints on sterile neutrinos and mirror photon}

The Big-Bang Nucleosynthesis (BBN) stringently restricts the
existence of the new relativistic degrees of freedom. The constraint
on the new degrees of freedom is conventionally quoted as the
effective number of additional light neutrinos. The latest Planck
2013 results show $N_{eff}^{}=3.30\pm 0.27$ \cite{ade2013}. So, one
light sterile neutrino can be allowed at $3\,\sigma$ level. A very
recent analysis \cite{afghm2013,kmms2013} on the neutrino
oscillation data also hint at the existence of an additional
neutrino with an eV-scale mass. This means the other two sterile
neutrinos should have the masses heavier than a few MeV and have a
lifetime shorter than $1$ second. From Eqs. (\ref{asmixing}) and
(\ref{mndecay}), we hence can put
\begin{eqnarray}
m_{\nu'}^{\textrm{max}}&>&
92\,\textrm{MeV}\left[\frac{(\langle\phi^{}_u\rangle/\langle\phi'^{}_u\rangle)^{2}_{}}{(U_{\nu\nu'}^\dagger
U_{\nu\nu'}^{})_{ii}^{}}\right]^{\frac{1}{5}}_{}
\left(\frac{\langle\phi'^{}_u\rangle/\langle\phi^{}_u\rangle}{2000}\right)^{\frac{2}{5}}_{}\nonumber\\
&&\times
\left(\frac{1\,\textrm{sec}}{\tau_{\nu'^{}_i}^{}}\right)^{\frac{1}{5}}_{}~~\textrm{for}
~~\tau_{\nu'^{}_i}^{}<1\,\textrm{sec}\,.
\end{eqnarray}
The mirror photon should also satisfy the BBN constraint. From Eq.
(\ref{mphoton}), it is easy to see
\begin{eqnarray}
\tau_{A'}^{}&\simeq&\left(\frac{4\times
10^{-11}_{}}{\epsilon}\right)^2_{}
\left(\frac{100\,\textrm{MeV}}{m_{A'}^{}}\right)\textrm{sec}\,.
\end{eqnarray}
Clearly, the mirror photon $A'$ with a mass
$m_{A'}^{}=100\,\textrm{MeV}$ can have a lifetime shorter than $1$
second if we take $\epsilon> 4\times 10^{-11}_{}$. Currently, the
measurement on the muon magnetic moment constrains $\epsilon^2
\cos^2_{}\theta^{}_W \cos^2_{}\theta'^{}_W< 2\times 10^{-5}_{}$ for
$m_{A'}^{}=100\,\textrm{MeV}$ \cite{pospelov2008}.

Furthermore, the active-sterile neutrino mass matrix
(\ref{asmatrixm}) should be constrained by the neutrinoless double
beta decay experiments. In the regime of
$m_{\nu'}^{\textrm{max}}\lesssim 100\,\textrm{MeV}$, we can perform
\cite{rodejohann2011}
\begin{eqnarray}
\label{nuless}
&&|m_{\beta\beta}^{}|=\left|\left(-h\frac{\langle\phi^{}_u\rangle^2_{}}{M_N^{}}h^T_{}\right)_{11}\right|<
0.2\,\textrm{eV}\nonumber\\
&&\Rightarrow
\left|\left(m_{\nu'}^{}\right)_{11}^{}\right|=\left|\sum_i^{}\left[\left(U_{\nu'}^{}\right)_{1i}^{}\right]^2_{}m_{\nu'^{}_i}^{}\right|\nonumber\\
&&\quad\quad\quad\quad\quad< 0.8\,\textrm{MeV}
\left(\frac{\langle\phi'^{}_u\rangle/\langle\phi^{}_u\rangle}{2000}\right)^2_{}\,.
\end{eqnarray}

\subsection{Leptogenesis scale}

From Eqs. (\ref{cpamax}) and (\ref{basymmetry}), the CP asymmetry
$|\varepsilon_{N_1^{}}^{}|$ should be bigger than
\begin{eqnarray}
\varepsilon_{N_1^{}}^{\textrm{max}}>5.45\times 10^{-8}_{}\,,
\end{eqnarray}
to explain the observed baryon asymmetry. Accordingly, we can have a
low limit on the leptogensis scale,
\begin{eqnarray}
M_{N_1^{}}^{}&>&1.3\times
10^{6}_{}\,\textrm{GeV}\left(\frac{100\,\textrm{MeV}}{m_{\nu'}^{\textrm{max}}}\right)
\left(\frac{\langle\phi'^{}_u\rangle/\langle\phi^{}_u\rangle}{2000}\right)^{2}_{}\,,\quad~~
\end{eqnarray}
which is expected to be below the critical temperature
(\ref{srate}),
\begin{eqnarray}
T_D^{}&>&1.1\times
10^{7}_{}\,\textrm{GeV}\left[\frac{2(m_{\nu'}^{\textrm{max}})^2_{}}{\sum_i^{}m_{\nu'^{}_i}^2}\right]
\left(\frac{100\,\textrm{MeV}}{m_{\nu'}^{\textrm{max}}}\right)^2_{}\nonumber\\
&&\times
\left[\frac{(\langle\phi'\rangle/\langle\phi\rangle)}{2000}\right]^{4}_{}\,.
\end{eqnarray}

\section{An example of parameter choice}

As an example, let us set
\begin{eqnarray}
\label{mewvev}
\langle\phi'^{}_{u}\rangle=2000\,\langle\phi^{}_{u}\rangle\,,~\tan\beta'=380\,,~\tan\beta=50\,,
\end{eqnarray}
to give the mirror charged fermion masses \cite{nakamura2010},
\begin{eqnarray}
\begin{array}{lcrclcr}
m_{d'}^{}&=&1.3\,\textrm{GeV}&\textrm{for}&m_{d}^{}&=&4.8\,\textrm{MeV}\,,\\
m_{u'}^{}&=&4.6\,\textrm{GeV}&\textrm{for}&m_{u}^{}&=&2.3\,\textrm{MeV}\,,\\
m_{s'}^{}&=&25\,\textrm{GeV}&\textrm{for}&m_s^{}&=&95\,\textrm{MeV}\,,\\
m_{c'}^{}&=&2.550\times 10^{3}_{}\,\textrm{TeV}&\textrm{for}&m_{c}^{}&=&1.275\,\textrm{GeV}\,,\\
m_{b'}^{}&=&1.10\times 10^3_{}\,\textrm{TeV}&\textrm{for}&m_{b}^{}&=&4.18\,\textrm{GeV}\,,\\
m_{t'}^{}&=&3.470\times 10^{5}\,\textrm{TeV}&\textrm{for}&m_{t}^{}&=&173.5\,\textrm{GeV}\,,\\
m_{e'}^{}&=&134\,\textrm{MeV}&\textrm{for}&m_{\mu}^{}&=&0.511\,\textrm{MeV}\,,\\
m_{\mu'}^{}&=&27.82\,\textrm{GeV}&\textrm{for}&m_{\mu}^{}&=&105.7\,\textrm{MeV}\,,\\
m_{\tau'}^{}&=&467.6\,\textrm{GeV}&\textrm{for}&m_{\tau}^{}&=&1.777\,\textrm{GeV}\,.
\end{array}
\end{eqnarray}
We hence can determine the mirror hadronic scale (\ref{mqcd}),
\begin{eqnarray}
\Lambda^{}_{\textrm{QCD}'}=1.28\,\textrm{GeV}~~\textrm{for}~~\Lambda_{\textrm{QCD}}^{}=200\,\textrm{MeV}\,.
\end{eqnarray}
and then the mirror baryon masses (\ref{mpnmass}) and
(\ref{mbmass}),
\begin{eqnarray}
m_{p'}^{}=10.5\,\textrm{GeV}\,,
~m_{n'}^{}=7.2\,\textrm{GeV}\,,~m_{\Delta'^{-}_{}}^{}=5\,\textrm{GeV}\,.
\end{eqnarray}
So, the mirror $\Delta'^{-}_{}$ baryon is the dark matter particle.
For the other parameter choice, the mirror proton or neutron can act
as the dark matter particle.

By further taking the masses of the sterile neutrinos,
\begin{eqnarray}
\label{spectrum}
\hat{m}_{\nu'}^{}&=&\textrm{diag}\{0.96\,\textrm{eV},~95\,\textrm{MeV},~100\,\textrm{MeV}\}\,,
\end{eqnarray}
as well as the VEV and Yukawa couplings of the Higgs bidoublet,
\begin{eqnarray}
\langle\Sigma\rangle=610\,\textrm{eV}~~\textrm{for}~~M_{\Sigma}^{}=10\,\rho=10^{13}_{}\,\textrm{GeV}\,,\quad\quad\quad&&\\
[2mm] f=\!\!\left[\begin{array}{rrr}i\,1.09\times
10^{-4}_{}&i\,1.87\times
10^{-4}_{}&-2.82\times 10^{-3}_{}\\
i\,1.87\times 10^{-4}_{}&-5.26\times 10^{-2}_{}&-3.46\times 10^{-2}_{}\\
-2.82\times 10^{-3}_{}&-3.46\times 10^{-2}_{}&-3.85\times 10^{-2}_{}
\end{array}\right],&&
\end{eqnarray}
the active neutrino masses (\ref{active}) can arrive at
\begin{eqnarray}
(m_{\nu}^{})_{ij}^{}&\simeq&
-f_{i1}^{}f_{j1}^{}\frac{\langle\Sigma\rangle^2_{}}{m_{\nu'^{}_1}^{}}-2f_{ij}
\frac{\langle\Sigma\rangle\langle\phi^{}_{u}\rangle}{\langle\phi'^{}_{u}\rangle}\nonumber\\
&\simeq&\textrm{eV}\left[\begin{array}{lll}0.00465&0.00793&
0.00172\\
0.00793&0.0321&0.0211\\
0.00172 & 0.0211 & 0.0235\end{array}\right]\,,
\end{eqnarray}
to give the mass eigenvalues and mixing angles,
\begin{eqnarray}
&&m_{\nu_1^{}}^{}\simeq 0.001\,\textrm{eV}\,,\nonumber\\
&&m_{\nu_2^{}}^2-m_{\nu_1^{}}^2\simeq 7.6\times
10^{-5}_{}\,\textrm{eV}^2_{}\,,\nonumber\\
&&m_{\nu_3^{}}^2-m_{\nu_1^{}}^2\simeq 2.55\times
10^{-3}_{}\,\textrm{eV}^2_{}\,,\nonumber\\
&&\sin^2_{}\theta_{12}^{}\simeq
0.32\,,\,\sin^2_{}\theta_{23}^{}\simeq 0.6\,,
\,\sin^2_{}\theta_{13}^{}\simeq 0.025\,,\nonumber\\
&&
\end{eqnarray}
which is consistent to the neutrino oscillation data.

The lightest sterile neutrino $\nu'^{}_1$ can help us to fit the
short baseline neutrino oscillation data if it has a mixing with the
active neutrinos as below \cite{kmms2013},
\begin{eqnarray}
|(U_{\nu\nu'}^{})_{11}^{}|=0.15\,,~~|(U_{\nu\nu'}^{})_{21}|=0.17\,.
\end{eqnarray}
This can be achieved by inputting
\begin{eqnarray}
\label{sbs} |(fU^{\ast}_{\nu'})_{11}^{}|\simeq2.36\times 10^{-4}\,,~
|(fU^{\ast}_{\nu'})_{21}^{}|\simeq2.68\times 10^{-4}\,,
\end{eqnarray}
in Eq. (\ref{asmixing}). As shown in Eq. (\ref{nuless}), the sterile
neutrino masses (\ref{spectrum}) will also be constrained by the
neutrinoless double beta decay experiments,
\begin{eqnarray}
\label{0nu} &&\left|9.6\times
10^{-9}\,[(U_{\nu'}^{})_{11}^{}]^2_{}+0.95\,
[(U_{\nu'}^{})_{12}^{}]^2_{}+[(U_{\nu'}^{})_{13}^{}]^2_{}\right|\nonumber\\
&&<0.008\,,
\end{eqnarray}
To fulfill the the constraints (\ref{sbs}) and (\ref{0nu}), we can
choose the mirror PMNS matrix $U_{\nu'}^{}$ to have the zero CP
phases and the following mixing angles,
\begin{eqnarray}
\sin\theta'^{}_{13}=0\,,~\sin\theta'^{}_{12}\simeq
0.0897\,,~\sin\theta'^{}_{23}\simeq 0.817\,,
\end{eqnarray}
and then give the required values of the elements
\begin{eqnarray}
(U_{\nu'}^{})_{11}=\cos\theta'^{}_{12}\,,~(U_{\nu'}^{})_{12}=\sin\theta'^{}_{12}\,,
~(U_{\nu'}^{})_{13}=0\,,~~&&\nonumber\\
[2mm] (U_{\nu'}^{})_{21}=-\sin\theta'^{}_{12}\cos\theta'^{}_{23}\,,~
(U_{\nu'}^{})_{31}=\sin\theta'^{}_{12}\sin\theta'^{}_{23}\,.&&
\end{eqnarray}

The Yukawa couplings in the sterile neutrino masses (\ref{sterile})
can be parameterized by \cite{ci2000}
\begin{eqnarray}
h=-\frac{i}{\langle\phi'^{}_u\rangle}U_{\nu'}^{}\sqrt{\hat{m}_{\nu'}^{}}\Omega^T_{}\sqrt{M_N^{}}\,,
\end{eqnarray}
with $\Omega$ being an arbitrary orthogonal matrix. By taking the
masses of the fermion singlets,
\begin{eqnarray}
M_{N}^{}=\textrm{diag}\{10^{7}_{}\,\textrm{GeV},
10^{8}_{}\,\textrm{GeV},10^{9}_{}\,\textrm{GeV}\}\,,
\end{eqnarray}
we can obtain the Yukawa couplings,
\begin{eqnarray}
h=-iU^{}_{\nu'}\quad\quad\quad\quad\quad\quad\quad\quad\quad\quad\quad\quad\quad\quad\quad\quad\quad\quad&&\nonumber\\
[2mm] \times\!\!\left[\!\begin{array}{ccc}2.82\cdot
10^{-7}_{}\,\Omega_{11}^{}&8.90\cdot 10^{-7}_{}\,\Omega_{21}
&2.82\cdot
10^{-6}_{}\,\Omega_{31}^{}\\
[2mm] 2.80\cdot 10^{-3}_{}\,\Omega_{12}^{}&8.86\cdot
10^{-3}_{}\,\Omega_{22}^{}
&2.80\cdot 10^{-2}_{}\,\Omega_{32}^{}\\
[2mm] 2.87\cdot 10^{-3}_{}\,\Omega_{13}^{}&9.09\cdot
10^{-3}_{}\,\Omega_{23}^{} &2.87\cdot
10^{-2}_{}\,\Omega_{33}^{}\end{array}\!\right].&&
\end{eqnarray}
By further
\begin{eqnarray}
|\Omega_{11}^{}|\,,~|\Omega_{21}^{}|\ll 1
\,,~~|\Omega_{31}^{}|\simeq 1\,,
\end{eqnarray}
we can get the out-of-equilibrium parameter (\ref{equilibrium}),
\begin{eqnarray}
K_{N_1^{}}^{}\simeq 0.016 \,,
\end{eqnarray}
and the CP asymmetry (\ref{scpa}),
\begin{eqnarray}
\varepsilon_{N_1^{}}^{}&\simeq&8.21\times
10^{-7}_{}\,\textrm{Re}(\Omega_{31}^{})\textrm{Im}(\Omega_{31}^{})\,.
\end{eqnarray}
So, the final baryon asymmetry (\ref{basymmetry}) can explain the
observation \cite{komatsu2010},
\begin{eqnarray}
\eta^{}_B=\eta'^{}_B\simeq 0.888\times
10^{-10}_{}~~\textrm{for}~~\varepsilon_{N_1^{}}^{}\simeq -5.45\times
10^{-8}_{}\,.
\end{eqnarray}

\section{Conclusion}

In this paper we have demonstrated a new mirror universe model,
which contains three gauge-singlet Majorana fermions and an
$[SU(2)^{}_L\times SU(2)'^{}_L]$-bidoublet Higgs scalar in addition
to the $SU(3)^{}_c\times SU(2)^{}_L\times U(1)^{}_Y$ ordinary sector
and its $SU(3)'^{}_c\times SU(2)'^{}_L\times U(1)'^{}_Y$ mirror
partner. In our model, the mirror sterile neutrino masses can have a
form of the canonical seesaw, while the ordinary active neutrino
masses can have a form of the double and linear seesaw. The mixing
between the active and sterile neutrinos can also be
seesaw-suppressed. Two sterile neutrinos can be above the MeV scale
to avoid the BBN constraint, while the third sterile neutrino can be
at the eV scale to fit the short baseline neutrino oscillation data.
An ordinary lepton asymmetry and an equal mirror lepton asymmetry
can be simultaneously produced from the decays of the fermion
singlets. The baryonic and dark matter asymmetries then can equal
each other since the ordinary and mirror sphaleron processes have a
same efficiency of lepton-to-baryon conversion. Consequently, the
lightest mirror baryon should have a mass around $5\,\textrm{GeV}$
to serve as the dark matter particle. The $U(1)^{}_Y$ and
$U(1)'^{}_Y$ kinetic mixing can mediate a testable dark matter
scattering.


\begin{thebibliography}{99}

\bibitem{ftv2012}
D.V. Forero, M. Tortola, and J.W.F. Valle, Phys. Rev. D \textbf{86},
073012 (2012); G.L. Fogli, E. Lisi, A. Marrone, D. Montanino, A.
Palazzo, and A.M. Rotunno, Phys. Rev. D \textbf{86}, 013012 (2012).




\bibitem{aguilar-arevalo2001}
A. Aguilar-Arevalo {\it et al.}, (LSND Collaboration), Phys. Rev. D
\textbf{64}, 112007 (2001); A.A. Aguilar-Arevalo {\it et al.},
(MiniBooNE Collaboration), Phys. Rev. Lett. \textbf{105}, 181801
(2010).

\bibitem{mueller2011}
Th.A. Mueller {\it et al.}, Phys. Rev. C \textbf{83}, 054615 (2011).



\bibitem{kms2011}
J. Kopp, M. Maltoni, and T. Schwetz, Phys. Rev. Lett. \textbf{107},
091801 (2011); C. Giunti and M. Laveder, Phys. Rev. D \textbf{84},
073008 (2011); C. Giunti and M. Laveder, Phys. Rev. D \textbf{84},
093006 (2011); C. Giunti and M. Laveder, Phys. Lett. B \textbf{706},
200 (2011); G. Karagiorgi, M.H. Shaevitz, and J.M. Conrad,
arXiv:1202.1024 [hep-ph]; A. Donini, P. Hernandez, J. Lopez-Pavon,
M. Maltoni, and T. Schwetz, JHEP \textbf{1207}, 161 (2012).



\bibitem{afghm2013}
A. Archidiacono, N. Fornengo, C. Giunti, S. Hannestad, and A.
Melchiorri, arXiv:1302.6720 [astro-ph.CO].


\bibitem{kmms2013}
J. Kopp, P.A.N. Machado, M. Maltoni, and T. Schwetz, arXiv:1303.3011
[hep-ph].


\bibitem{minkowski1977}
P. Minkowski, Phys. Lett. B \textbf{67}, 421 (1977); T. Yanagida, in
{\it Proceedings of the Workshop on Unified Theory and the Baryon
Number of the Universe}, edited by O. Sawada and A. Sugamoto (KEK,
Tsukuba, 1979), p. 95; M. Gell-Mann, P. Ramond, and R. Slansky, in
{\it Supergravity}, edited by F. van Nieuwenhuizen and D. Freedman
(North Holland, Amsterdam, 1979), p. 315; S.L. Glashow, in {\it
Quarks and Leptons}, edited by M. L\'{e}vy {\it et al.} (Plenum, New
York, 1980), p. 707; R.N. Mohapatra and G. Senjanovi\'{c}, Phys.
Rev. Lett. \textbf{44}, 912 (1980).


\bibitem{mw1980}
M. Magg and C. Wetterich, Phys. Lett. B \textbf{94}, 61 (1980); J.
Schechter and J.W.F. Valle, Phys. Rev. D \textbf{22}, 2227 (1980);
T.P. Cheng and L.F. Li, Phys. Rev. D \textbf{22}, 2860 (1980); G.
Lazarides, Q. Shafi, and C. Wetterich, Nucl. Phys. B \textbf{181},
287 (1981); R.N. Mohapatra and G. Senjanovi\'{c}, Phys. Rev. D
\textbf{23}, 165 (1981).


\bibitem{flhj1989}
R. Foot, H. Lew, X.G. He, and G.C. Joshi, Z. Phys. C \textbf{44},
441 (1989).


\bibitem{zee1980}
A. Zee, Phys. Lett. B \textbf{93}, 389 (1980); A. Zee, Phys. Lett. B
\textbf{161}, 141 (1985); K.S. Babu, Phys. Lett. B \textbf{203}, 132
(1988);



\bibitem{mahanta1999}
U. Mahanta, Phys. Rev. D \textbf{62}, 073009 (2000); K.S. Babu and
C.N. Leung, Nucl. Phys. B \textbf{619}, 667 (2001); A. de Gouv\^{e}a
and J. Jenkins, Phys. Rev. D \textbf{77}, 013008 (2008); D.
Aristizabal Sierra, M. Hirsch, and S.G. Kovalenko, Phys. Rev. D
\textrm{77}, 055011 (2008); K.S. Babu and J. Julio, Nucl. Phys. B
\textbf{841}, 130 (2010); K.S. Babu and J. Julio, Phys. Rev. D
\textbf{85}, 073005 (2012); M. Kohda, H. Sugiyama, and K. Tsumura,
Phys. Lett. B \textbf{718}, 1436 (2013); C.S. Chen and L.H. Tsai,
arXiv:1210.6264 [hep-ph].

\bibitem{knt2002}
L.M. Krauss, S. Nasri, and M. Trodden, Phys. Rev. D \textbf{67},
085002 (2003); K. Cheung and O. Seto, Phys. Rev. D \textbf{69},
113009 (2004); E. Ma, Phys. Rev. D \textbf{73}, 077301 (2006); P.H.
Gu, arXiv:1203.4165 [hep-ph]; M. Aoki, J. Kubo, and H. Takano,
arXiv:1302.3936 [hep-ph].




\bibitem{gs2008}
P.H. Gu and U. Sarkar, Phys. Rev. D \textbf{77}, 105031 (2008); M.K.
Parida, Phys. Lett. B \textbf{704}, 206 (2011).


\bibitem{koy2013}
Y. Kajiyama, H. Okada and K. Yagyu, arXiv:1303.3463 [hep-ph].



\bibitem{rw1983}
M. Roncadelli and D. Wyler, Phys. Lett. B \textbf{133}, 325 (1983);
P. Roy and O. Shanker, Phys. Rev. Lett. \textbf{52}, 713 (1984); H.
Murayama and A. Pierce, Phys. Rev. Lett. \textbf{89}, 271601 (2002);
P.H. Gu and H.J. He, JCAP \textbf{0612}, 010 (2006).


\bibitem{gs2007}
P.H. Gu and U. Sarkar, Phys. Rev. D \textbf{77}, 105031 (2008); S.
Kanemura, T. Nabeshima, and H. Sugiyama, Phys. Rev. D \textbf{85},
033004 (2012); S. Kanemura, T. Nabeshima, and H. Sugiyama, Phys.
Rev. D \textbf{87}, 015009 (2013); Y. Farzan and E. Ma, Phys. Rev. D
\textbf{86}, 033007 (2012).

\bibitem{mohapatra1986}
R.N. Mohapatra, Phys. Rev. Lett. \textbf{56}, 561 (1986).




\bibitem{barr2003}
S.M. Barr, Phys. Rev. Lett. \textbf{92}, 101601 (2004).




\bibitem{mv1986}
R.N. Mohapatra and J.W.F. Valle, Phys. Rev. D \textbf{34}, 1642
(1986); M.C. Gonzalez-Garcia and J.W.F. Valle, Phys. Lett. B
\textbf{216}, 360 (1989).


\bibitem{fp2009}
P. Fileviez Perez, JHEP \textbf{0903}, 142 (2009); J. Chakrabortty,
S. Goswami, and A. Raychaudhuri, Phys. Lett. B \textbf{698}, 265
(2011).

\bibitem{gu2011}
P.H. Gu, Phys. Rev. D \textrm{84}, 097301 (2011).







\bibitem{ly1956}
T.D. Lee and C.N. Yang, Phys. Rev. textbf{104}, 254 (1956).


\bibitem{kop1966}
I.Yu. Kobzarev, L.B. Okun, and I.Ya. Pomeranchuk, Sov. J. Nucl.
Phys. \textbf{3}, 837 (1966) [Yad. Fiz. \textbf{3}, 1154 (1966)].

\bibitem{pavsic1974}
M. Pavsic, Int. J. Theor. Phys. \textbf{9}, 229 (1974).


\bibitem{bk1982}
S.I. Blinnikov and M.Y. Khlopov, Sov. J. Nucl. Phys. \textbf{36},
472 (1982) [Yad. Fiz. \textbf{36}, 809 (1982)]; S.I. Blinnikov and
M.Y. Khlopov, Sov. Astron. \textbf{27}, 371 (1983) [Astro. Zh.
\textbf{60}, 632 (1983)].


\bibitem{glashow1986}
S. L. Glashow, Phys. Lett. B \textbf{167}, 35 (1986).


\bibitem{flv1991}
R. Foot, H. Lew, and R.R. Volkas, Phys. Lett. B \textbf{272}, 67
(1991).







\bibitem{silagadze1997}
Z. Silagadze, Phys. Atom. Nucl. \textbf{60}, 272 (1997) [Yad. Fiz.
\textbf{60N2}, 336 (1997)].





\bibitem{hodges1993}
H.M. Hodges, Phys. Rev. D \textbf{47} 456 (1993).





\bibitem{cf1998}
M. Collie and R. Foot, Phys. Lett. B \textbf{432}, 134 (1998); R.
Foot and R.R. Volkas, Phys. Rev. D \textbf{61}, 043507 (2000).



\bibitem{mt1999}
R.N. Mohapatra and V.L. Teplitz, Phys. Lett. B \textbf{462}, 302
(1999); R.N. Mohapatra and V.L. Teplitz, Phys. Rev. D \textbf{62},
063506 (2000); R.N. Mohapatra, S. Nussinov, and V.L. Teplitz, Phys.
Rev. D \textbf{66}, 063002 (2002).






\bibitem{bcv2001}
Z. Berezhiani, D. Comelli, and F.L. Villante, Phys. Lett. B
\textbf{503} 362 (2001); Z. Berezhiani, P. Ciarcelluti, D. Comelli,
and F.L. Villante, Int. J. Mod. Phys. D 14, 107 (2005); P.
Ciarcelluti, Int. J. Mod. Phys. D 14, 187 (2005).




\bibitem{bgg2001}
Z. Berezhiani, L. Gianfagna, and M. Giannotti, Phys. Lett. B
\textbf{500}, 286 (2001); Z. Berezhiani and A. Lepidi, Phys. Lett. B
\textbf{681}, 276 (2009).



\bibitem{iv2003}
A.Y. Ignatiev and R.R. Volkas, Phys. Rev. D \textbf{68}, 023518
(2003).






\bibitem{fv2003}
R. Foot and R.R. Volkas, Phys. Rev. D \textbf{68}, 021304 (2003); R.
Foot and R.R. Volkas, Phys. Rev. D \textbf{69}, 123510 (2004).








\bibitem{foot2004}
R. Foot, Phys. Rev. D \textbf{69}, 036001 (2004); R. Foot, Phys.
Rev. D \textbf{74}, 023514 (2006); R. Foot, Phys. Rev. D
\textbf{78}, 043529 (2008); P. Ciarcelluti and R. Foot, Phys. Lett.
B \textbf{679}, 278 (2009); R. Foot, Phys. Rev. D \textbf{81},
087302 (2010); R. Foot, Phys. Rev. D \textbf{86}, 023524 (2012);
Phys. Lett. B \textbf{718}, 745 (2013).



\bibitem{acmy2009}
H. An, S.L. Chen, R.N. Mohapatra, and Y. Zhang, JHEP \textbf{1003},
124 (2010); H. An, S.L. Chen, R.N. Mohapatra, S. Nussinov, and Y.
Zhang, Phys. Rev. D \textbf{82}, 023533 (2010).


\bibitem{dlnt2011}
C.R. Das, L.V. Laperashvili, H.B. Nielsen, and A. Tureanu, Phys.
Rev. D \textbf{84}, 063510 (2011).

\bibitem{chly2012}
J.W. Cui, H.J. He, L.C. L\"{u}, and F.R. Yin, Phys. Rev. D
\textbf{85}, 096003 (2012).






\bibitem{gu2012-2}
P.H. Gu, arXiv:1209.4579 [hep-ph].





\bibitem{foot2012}
R. Foot, arXiv:1211.1500 [astro-ph.CO]; R. Foot, arXiv:1303.1727
[astro-ph.GA].





\bibitem{flv1992}
R. Foot, H. Lew, and R.R. Volkas, Mod. Phys. Lett. A \textbf{7},
2567 (1992); R. Foot and R.R. Volkas, Phys. Rev. D \textbf{52}, 6595
(1995).




\bibitem{abs1992}
E.H. Akhmedov, Z. Berezhiani, and G. Senjanovi\'{c}, Phys. Rev.
Lett. \textbf{69}, 3013 (1992); Z. Berezhiani and R.N. Mohapatra,
Phys. Rev. D \textbf{52}, 6607 (1995).




\bibitem{bdm1996}
Z. Berezhiani, Acta. Phys. Polon. B \textbf{27}, 1503 (1996); Z.
Berezhiani, A. Dolgov, and R.N. Mohapatra, Phys. Lett. B
\textbf{375}, 26 (1996).


\bibitem{bb2001}
L. Bento and Z. Berezhiani, Phys. Rev. Lett. \textbf{87}, 231404
(2001).

\bibitem{bb2001-2}
L. Bento and Z. Berezhiani, Phys. Rev. D \textbf{64}, 115015 (2001).


\bibitem{berezhiani2004}
L. Bento and Z. Berezhiani, Int. J. Mod. Phys. A \textbf{19}, 3775
(2004)


\bibitem{berezhiani2005}
Z. Berezhiani, hep-ph/0508233.

\bibitem{berezhiani2006}
Z. Berezhiani, AIP Conf. Proc. \textbf{878}, 195 (2006); Z.
Berezhiani, Eur. Phys. J. ST \textbf{163}, 271 (2008).




\bibitem{bb2006-1}
L. Bento and Z. Berezhiani, Phys. Rev. Lett. \textbf{96}, 081801
(2006).

\bibitem{bb2006}
L. Bento and Z. Berezhiani, Phys. Lett. B \textbf{635}, 253 (2006).




\bibitem{kty2010}
A. Kusenko, F. Takahashi, and T.T. Yanagida, Phys. Lett. B
\textbf{693}, 144 (2010); A. Adulpravitchai, and R. Takahashi, JHEP
\textbf{1109}, 127 (2011).


\bibitem{mohapatra1991}
R.N. Mohapatra, Phys. Rev. D \textbf{64}, 091301 (2001); M.
Shaposhnikov, Nucl. Phys. B \textbf{763}, 49 (2007); M. Lindner, A.
Merle, and V. Niro, JCAP \textbf{1101}, 034 (2011).


\bibitem{gu2010}
P.H. Gu, Phys. Rev. D \textbf{82}, 093009 (2010); P.H. Gu, Phys.
Rev. D \textbf{81}, 095002 (2010).


\bibitem{mn2011}
A. Merle and V. Niro, JCAP \textbf{1107}, 023 (2011); J. Barry, W.
Rodejohann, and H. Zhang, JHEP \textbf{1107}, 091 (2011); J. Barry,
W. Rodejohann, and H. Zhang, JHEP \textbf{1201}, 052 (2012).


\bibitem{gh2011}
A. de Gouv\^{e}a and W.C. Huang, Phys. Rev. D \textbf{85}, 053006
(2012); J. Fan and P. Langacker, JHEP \textbf{1204}, 083 (2012).

\bibitem{ct2011}
C.S. Chen and R. Takahashi, Eur. Phys. J. C \textbf{72}, 2089
(2012).



\bibitem{abazajian2012}
For a recent view, see K.N. Abazajian {\it et al.}, arXiv:1204.5379
[hep-ph].






\bibitem{komatsu2010}
E. Komatsu {\it et al.}, Astrophys. J. Suppl. \textbf{192}, 18
(2011).





\bibitem{nussinov1985}
S. Nussinov, Phys. Lett. B \textbf{165}, 55 (1985).

\bibitem{bcf1990}
S.M. Barr, R.S. Chivulula, and E. Farhi, Phys. Lett. B \textbf{241},
387 (1990); S.M. Barr, Phys. Rev. D \textbf{44}, 3062 (1991).

\bibitem{kaplan1992}
D.B. Kaplan, Phys. Rev. Lett. \textbf{68}, 741 (1992).


\bibitem{dgw1992}
S. Dodelson, B.R. Greene, and L.M. Widrow, Nucl. Phys. B
\textbf{372}, 467 (1992).

\bibitem{kuzmin1998}
V.A. Kuzmin, Phys. Part. Nucl. \textbf{29}, 257 (1998) [Fiz. Elem.
Chast. Atom. Yadra \textbf{29}, 637 (1998)].


\bibitem{kl2005}
R. Kitano and I. Low, Phys. Rev. D \textbf{71}, 023510 (2005).


\bibitem{as2005}
K. Agashe and G. Servant, JCAP \textbf{0502}, 002 (2005); M.
Cirelli, P. Panci, G. Servant, and G. Zaharijas, JCAP \textbf{1203},
015 (2012).



\bibitem{clt2005}
N. Cosme, L. Lopez Honorez, and M.H.G. Tytgat, Phys. Rev. D
\textbf{72}, 043505 (2005).







\bibitem{gsz2009}
P.H. Gu, U. Sarkar, and X. Zhang, Phys. Rev. D \textbf{80}, 076003
(2009); P.H. Gu and U. Sarkar, Phys. Rev. D \textbf{81}, 033001
(2010); P.H. Gu, M. Lindner, U. Sarkar, and X. Zhang, Phys. Rev. D
\textbf{83}, 055008 (2011).



\bibitem{dmst2010}
H. Davoudiasl, D.E. Morrissey, K. Sigurdson, and S. Tulin, Phys.
Rev. Lett. \textbf{105}, 211304 (2010); H. Davoudiasl, D.E.
Morrissey, K. Sigurdson, and S. Tulin, Phys. Rev. D \textbf{84},
096008 (2011); N. Blinov, D.E. Morrissey, and S. Tulin,
arXiv:1206.3304 [hep-ph].


\bibitem{bdfr2011}
M. Blennow, B. Dasgupta, E. Fernandez-Martinez, and N. Rius, JHEP
\textbf{1103}, 014 (2011); M. Blennow, E. Fernandez-Martinez, J.
Rendondo, and P. Serra, arXiv:1203.5805 [hep-ph].

\bibitem{mcdonald2011}
J. McDonald, Phys. Rev. D \textbf{83}, 083509 (2011); Phys. Rev. D
\textbf{84}, 103514 (2011).






\bibitem{hmw2010}
L.J. Hall, J. March-Russell, and S.M. West, arXiv:1010.0245
[hep-ph]; J. March-Russell, M. McCullough, JCAP \textbf{1203}, 019
(2012); J. March-Russell, J. Unwin, and S.M. West, arXiv:1203.4854
[hep-ph].


\bibitem{dk2011}
B. Dutta and J. Kumar, Phys. Lett. B \textbf{699}, 364 (2011).


\bibitem{frv2011}
A. Falkowski, J.T. Ruderman, and T. Volansky, JHEP \textbf{1105},
106 (2011).






\bibitem{hms2011}
N. Haba, S. Matsumoto, and R. Sato, Phys. Rev. D \textbf{84}, 055016
(2011).


\bibitem{kllly2011}
Z. Kang, J. Li, T. Li, T. Liu, and J. Yang, arXiv:1102.5644
[hep-ph]; Z. Kang and T. Li, arXiv:1111.7313 [hep-ph].


\bibitem{gsv2011}
M.L. Graesser, I.M. Shoemaker, and L. Vecchi, JHEP \textbf{1110},
110 (2011).






\bibitem{fss2011}
M.T. Frandsen, S. Sarkar, and K. Schmidt-Hoberg, Phys. Rev. D
\textbf{84}, 051703 (2011).

\bibitem{myz2012}
S.D. McDermott, H.B. Yu, and K.M. Zurek, Phys. Rev. D \textbf{85},
023519 (2012); S. Tulin, H.B. Yu, and K.M. Zurek, JCAP
\textbf{1205}, 013 (2012).

\bibitem{idc2011}
H. Iminniyaz, M. Drees, and X. Chen, JCAP \textbf{107}, 003 (2011).



\bibitem{bpsv2011}
N.F. Bell, K. Petraki, I.M. Shoemaker, and R.R. Volkas, Phys. Rev. D
\textbf{84}, 123505 (2011); K. Petraki, M. Trodden, and R.R. Volkas,
JCAP \textbf{1202}, 044 (2012); B. von Harling, K. Petraki, and R.R.
Volkas, JCAP \textbf{1205}, 021 (2012).



\bibitem{cr2011}
Y. Cui, L. Randall, and B. Shuve, JHEP \textbf{1108}, 073 (2011); Y.
Cui, L. Randall, and B. Shuve, JHEP \textbf{1204}, 075 (2012).

\bibitem{as2012}
C. Arina, and N. Sahu, Nucl. Phys. B \textbf{854}, 666 (2012); C.
Arina, J.O. Gong, and N. Sahu, arXiv: 1206.0009 [hep-ph].




\bibitem{ms2011}
E. Ma and U. Sarkar, arXiv:1111.5350 [hep-ph].



\bibitem{dm2012}
H. Davoudiasl and R.N. Mohapatra, arXiv:1203.1247 [hep-ph].

\bibitem{fnp2012}
W.Z. Feng, P. Nath, and G. Peim, arXiv:1204.5752 [hep-ph].






\bibitem{bp2011}
M.R. Buckley and S. Profumo, Phys. Rev. Lett. \textbf{108}, 011301
(2012).


\bibitem{ptv2011}
K. Petraki, M. Trodden, R.R. Volkas, JCAP \textbf{1202}, 044 (2012).


\bibitem{ams2012}
C. Arina, R.N. Mohapatra, and N. Sahu, Phys. Lett. B \textbf{720},
130 (2013).


\bibitem{kv2012}
H. Kuismanen and L. Vija, Phys. Rev. D \textbf{87}, 015005 (2013).

\bibitem{ccs2012}
K.Y. Choi, E.J. Chun, and C.S. Shin, arXiv:1211.5409 [hep-ph].


\bibitem{bfk2013}
J. Bramente and K. Fukushima, and J. Kumar, arXiv:1301.0036
[hep-ph].


\bibitem{bmp2013}
N.F. Bell, A. Melatos, and K. Petraki, arXiv:1301.6811 [hep-ph].



\bibitem{fy1986}
M. Fukugita and T. Yanagida, Phys. Lett. B \textbf{174}, 45 (1986).

\bibitem{mz1992}
R.N. Mohapatra and X. Zhang, Phys. Rev. D \textbf{46}, 5331 (1992);
E. Ma and U. Sarkar, Phys. Rev. Lett. \textbf{80}, 5716 (1998).

\bibitem{fps1995}
M. Flanz, E.A. Paschos, and U. Sarkar, Phys. Lett. B \textbf{345},
248 (1995); M. Flanz, E.A. Paschos, U. Sarkar, and J. Weiss, Phys.
Lett. B \textbf{389}, 693 (1996); L. Covi, E. Roulet, and F.
Vissani, Phys. Lett. B \textbf{384}, 169 (1996); A. Pilaftsis, Phys.
Rev. D \textbf{56}, 5431 (1997).

\bibitem{hms2000}
T. Hambye, E. Ma, and U. Sarkar, Nucl. Phys. B \textbf{602}, 23
(2001).


\bibitem{di2002}
S. Davidson and A. Ibarra, Phys. Lett. B \textbf{535}, 25 (2002); W.
Buchm\"{u}ller, P. Di Bari, and M. Pl\"{u}macher, Nucl. Phys. B
\textbf{665}, 445 (2003).


\bibitem{hs2004}
T. Hambye and G. Senjanovi\'{c}, Phys. Lett. B \textbf{582}, 73
(2004); S. Antusch and S.F. King, Phys. Lett. B \textbf{597}, 199
(2004).

\bibitem{hlnps2004}
T. Hambye, Y. Lin, A. Notari, M. Papucci, A. Strumia, Nucl. Phys. B
\textbf{695}, 169 (2004).


\bibitem{hrs2005}
T. Hambye, M. Raidal, and A. Strumia, Phys. Lett. B \textbf{632},
667 (2006).


\bibitem{dnn2008}
S. Davidson, E. Nardi, and Y. Nir, Phys. Rept. \textbf{466}, 105
(2008).


\bibitem{bd2009}
S. Blanchet and P. Di Bari, Nucl. Phys. B \textbf{807}, 155 (2009).




\bibitem{gu2012}
P.H. Gu, Phys. Lett. B \textbf{713}, 485 (2012).


\bibitem{krs1985}
V.A. Kuzmin, V.A. Rubakov, and M.E. Shaposhnikov, Phys. Lett. B
\textbf{155}, 36 (1985).



\bibitem{gk2007}
I.F. Ginzburg and K.A. Kanishev, Phys. Rev. D \textbf{76}, 095013
(2007).


\bibitem{bcrwy2009}
M. Baumgart, C. Cheung, J.T. Runderman, L.T. Wang, and I. Yavin,
JHEP \textbf{0904}, 014 (2009).



\bibitem{fh1991}
R. Foot and X.G. He, Phys. Lett. B \textbf{267}, 509 (1991).





\bibitem{nakamura2010}
K. Nakamura {\it et al.}, (Particle Data Group), J. Phys. G
\textbf{37}, 075021 (2010).


\bibitem{mns1962}
Z. Maki, M. Nakagawa, and S. Sakata, Prog. Theor. Phys. \textbf{28},
870 (1962); B. Pontecorvo, Sov. Phys. JETP \textbf{26}, 984 (1968),
Zh. Eksp. Teor. Fiz. \textbf{53}, 1717 (1967).






\bibitem{bhl2009}
F. Bezrukov, H. Hettmansperger, and M. Lindner, Phys. Rev. D
\textbf{81}, 085032 (2010).


\bibitem{kt1990}
E.W. Kolb and M.S. Turner, \textit{The Early Universe},
Addison-Wesley, 1990.

\bibitem{fy1990}
M. Fukugita and T. Yanagida, Phys. Rev. D \textbf{42}, 1285 (1990).






\bibitem{angle2011}
J. Angle {\it et al.}, (XENON10 Collaboration), Phys. Rev. Lett.
\textbf{107}, 051301 (2011).



\bibitem{ade2013}
P.A.R. Ade {\it et al.}, (Planck Collaboration), arXiv:1303.5076
[astro-ph.CO].



\bibitem{pospelov2008}
M. Pospelov, Phys. Rev. D \textbf{80}, 095002 (2009).


\bibitem{rodejohann2011}
For recent reviews, see W. Rodejohann, Int. J. Mod. Phys. E
\textbf{20}, 1833 (2011).


\bibitem{ci2000}
J.A. Casas and A. Ibarra, Nucl. Phys. B \textbf{618}, 171 (2001).



\end{thebibliography}
\end{document}